\documentclass[prb,twocolumn,superscriptaddress,%
preprintnumbers,amsmath,amssymb]{revtex4}
\usepackage{amsmath}
\usepackage{graphicx}
\usepackage{dcolumn}
\usepackage{color}
\DeclareGraphicsExtensions{.png .jpg .pdf}

\begin{document}

\title{Valley filtering using electrostatic potentials in bilayer graphene}

\author{D. R. da Costa}\email{diego_rabelo@fisica.ufc.br}
\affiliation{Departamento de F\'isica, Universidade
Federal do Cear\'a, Caixa Postal 6030, Campus do Pici, 60455-900
Fortaleza, Cear\'a, Brazil}\affiliation{Department of Physics, University of
Antwerp, Groenenborgerlaan 171, B-2020 Antwerp,
Belgium}
\author{Andrey Chaves}\email{andrey@fisica.ufc.br}
\author{S. H. R. Sena}
\author{G. A. Farias}\email{gil@fisica.ufc.br}
\affiliation{Departamento de F\'isica, Universidade
Federal do Cear\'a, Caixa Postal 6030, Campus do Pici, 60455-900
Fortaleza, Cear\'a, Brazil}
\author{F. M. Peeters}\affiliation{Department of Physics, University of
Antwerp, Groenenborgerlaan 171, B-2020 Antwerp,
Belgium}\affiliation{Departamento de F\'isica, Universidade
Federal do Cear\'a, Caixa Postal 6030, Campus do Pici, 60455-900
Fortaleza, Cear\'a, Brazil}

\date{ \today }

\begin{abstract}
Propagation of an electron wave packet through a quantum point contact (QPC) defined by electrostatic gates in bilayer graphene is investigated. The gates provide a bias between the layers, in order to produce an energy gap. If the gates on both sides of the contact produce the same bias, steps in the electron transmission probability are observed, as in the usual QPC. However, if the bias is inverted on one of the sides of the QPC, only electrons belonging to one of the Dirac valleys are allowed to pass, which provides a very efficient valley filtering.

\noindent PACS number(s): 81.05.U-, 73.63.-b, 72.80.Vp
\end{abstract}

\maketitle

\section{Introduction}

The unique band structure of graphene has brought the possibility of developing devices based on different degrees of freedom, other than charge (electronics) and spin (spintronics), namely, using its different pseudo-spin states (pseudo-spintronics) and electronic valleys (valleytronics). Valley filtering in graphene has been pursued by many researchers, as a path to use the valley degree of freedom of electrons in this material as the basis for future valley-tronics. Previous theoretical proposals for valley filtering demand a high control of the atomic structure of the graphene layer, either by cutting it in specific directions as to produce uniform zigzag edges \cite{Beenakker}, or by applying stress in a specific manner in order to obtain an almost uniform pseudo-magnetic field \cite{Tony, Tony1, Andrey3}, or even by taking advantage of the valley filtering process that occurs when an electron propagates through a line of heptagon-pentagon defects on the honeycomb lattice \cite{Gunlycke, Gunlycke1, Liu}.

Monolayer graphene is gapless and therefore it poses problems to use it in some electronic devices\cite{CastroNetoReview}. In bilayer graphene on the other hand, a gap may be opened by applying a bias between the two layers \cite{KoshinoReview}. Therefore, in bilayer graphene, it is possible to produce electrostatic confined structures, such as quantum wires, dots and rings \cite{Milton, Milton1, Szafran, Zarenia, Zarenia1, Leandro}. A special case of quantum wire confinement occurs when one applies opposite bias on the different sides of the quantum wire potential: in this case, one dimensional uni-directional chiral states are created, whose subband structures along the free direction for $K$ and $K'$ valleys are mirror symmetric \cite{Morpurgo, Schomerus, Zarenia2}. In the present paper we will use the latter property to propose a novel valley filter, which is solely based on the use of electrostatic potentials and we do not require any complicated tailoring of the graphene lattice as needed in previous proposed filters. Although previous works \cite{Morpurgo, Zarenia2} have already studied the energy bands and chirality of these one-dimensional states created in bilayer graphene by electrostatic lateral confinement structures, it is important to investigate how to optimize the valley polarization efficiency that comes from the valley asymmetry of the bilayer graphene band structure observed for the different cones $K$ and $K'$ in this system.

\begin{figure}[!bpht]
\centerline{\includegraphics[width = \linewidth]{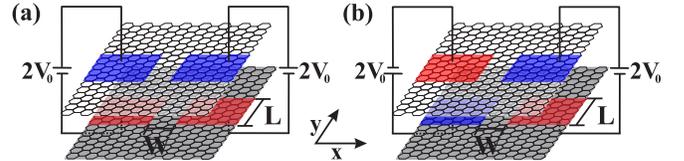}}
\caption{Sketch of the QPC structure, forming a channel with length $L$ and width $W$, with (a) aligned and (b) anti-aligned bias. The actual sample used in our numerical calculation is retangular with $3601\times 1000$ atoms in each layer that corresponds to a size $\approx 213\times443$ nm$^2$.} \label{fig:Fig1}
\end{figure}

In this paper, we demonstrate that a quantum point contact (QPC) defined by electrostatic gates in bilayer graphene, as sketched in Fig. \ref{fig:Fig1}(a), exhibits steps in its transmission probabilities as the energy of the incident electron increases, just like in an ordinary point contact. On the other hand, valley polarized current is predicted when the sides of the point contact have opposite bias, as sketched in Fig. \ref{fig:Fig1}(b). The specific conditions for such polarization are discussed in detail in Sec. \ref{sec.results}. In Sec. \ref{sec.technique} we present our technique to solve the time-dependent Schr\"odinger equation based on a tight-binding model. The numerical results are presented in Sec. \ref{sec.results} and we summarize our discussion giving the main conclusions in Sec. \ref{conclusion}.

\section{Split-operator technique for the bilayer graphene Hamiltonian}\label{sec.technique}

In this section we present the theoretical tools for the carrier time evolution in bilayer graphene. In order to do so, we solve the time-dependent Schr\"odinger equation for the tight-binding Hamiltonian of bilayer graphene in order to investigate the time evolution of a wave packet describing an electron propagating through a quantum point contact.

The time evolution of a quantum state is described by the time-dependent Schr\"odinger equation given by
\begin{equation}
\Psi(\vec{r},t) = \hat{U}(t,t_0) \Psi(\vec{r},t_0)
,\end{equation}
where $\hat{U}(t,t_0)$ is known as the time evolution operator. For the case in which the Hamiltonian does not explicitly depend on time, this operator can be written as $\hat{U}(t,t_0) = \exp\left[-\frac{i}{\hbar}H(t-t_0)\right]$. Different techniques to expand this exponential operator are found in the literature, for example iterative methods based on the Crank-Nicholson scheme and the Chebyschev polynomials method. Furthermore, for systems with moderate space dimensions there is the possibility to solve this problem by brute force, using full diagonalisation \cite{Fehske}. Here, we opted for the split-operator technique because of its elegance and its advantages: (i) solving time-dependent Schr\"odinger equation in this way is fast and much easier than using e. g. Green's functions techniques, (ii) results obtained by these approaches, while still physically meaningful and correct, are more pedagogical for the understanding of transport properties in quantum systems, as this method allows one, for instance, to track the center of mass trajectories, see reflection patterns and investigate valley-polarization by taking Fourier transform of the wave packet at each time step (this even makes it possible to study time-dependent valley scattering).\cite{Andrey3}

Our approach is based on the tight-binding model for the description of an electron in bilayer graphene. We consider respectively $n$ and $m$ as the row and column indexes to locate a particular site in the lattice, and $l = \{1, 2\}$ index corresponds to bottom and top layers, respectively. The basis vector state is defined as $|n,m,l\rangle$. So, the tight-binding Hamiltonian reads
\begin{align}\label{eq.HTB_BI}
& H_{TB}|n,m,l\rangle \cong (E_{n,m,l} + V_{n,m,l})|n,m,l\rangle\nonumber\\
& + \tau_{n-1,m}|n-1,m,l\rangle + \tau_{n+1,m}|n+1,m,l\rangle\nonumber\\
& + \tau_{n,m-1}|n,m-1,l\rangle + \tau_{n,m+1}|n,m+1,l\rangle\nonumber\\
& + \Delta_{n,m}|n,m,l+1\rangle + \Delta_{n,m}|n,m,l-1\rangle
,\end{align}
where $\tau_{n,m-1}$ and $\Delta_{n,m}$ are the intra- and inter-layer hopping energies between the sites, respectively. The tight-binding Hamiltonian for bilayer graphene in matrix form is now represented by two pentadiagonal matrices in blocks, connected by two diagonal matrices. To numerically simplify the problem, which is important when dealing with large systems, we first rewrite Eq. (\ref{eq.HTB_BI}) as follows
\begin{equation}
H_{TB}|n,m,l\rangle = H_{n,l}|n,m,l\rangle + H_{m,l}|n,m,l\rangle + H_{n,m}|n,m,l\rangle
,\end{equation}
where the operators $H_{n,l}$, $H_{m,l}$ and $H_{n,m}$ are defined as
\begin{align}
& H_{n,l}|n,m,l\rangle = \left(\frac{\epsilon_{n,m,l} + V_{n,m,l}}{2}\right)|n,m,l\rangle\nonumber\\
& + \tau_{n,m-1}|n,m-1,l\rangle + \tau_{n,m+1}|n,m+1,l\rangle
,\end{align}
\begin{align}
& H_{m,l}|n,m,l\rangle = \left(\frac{\epsilon_{n,m,l} + V_{n,m,l}}{2}\right)|n,m,l\rangle\nonumber\\
& + \tau_{n-1,m}|n-1,m,l\rangle + \tau_{n+1,m}|n+1,m,l\rangle
\end{align}
and
\begin{align}
H_{n,m}|n,m,l\rangle = \Delta_{n,m}|n,m,l+1\rangle + \Delta_{n,m}|n,m,l-1\rangle
.\end{align}

In doing so, we split the Hamiltonian and thus transform the problem of pentadiagonal matrices in blocks into a series of calculations involving only products of tridiagonal matrices, which are much easier to handle with known computational routines.

Subsequently, the time evolution operator is expanded as follows
\begin{align}\label{eq.TimeEvolutionOperatorTB}
e^{-(i/\hbar)H_{TB}\Delta t} &= e^{-(i/2\hbar)H_{n,m} \Delta t} e^{-(i/2\hbar)H_{m,l} \Delta t} e^{-(i/\hbar)H_{n,l}\Delta t}\nonumber\\
&\times e^{-(i/2\hbar)H_{m,l} \Delta t} e^{-(i/2\hbar)H_{n,m} \Delta t} + {\cal O}(\Delta t^3)
,\end{align}
and we neglect terms of order ${\cal O}(\Delta t^3)$ which correspond to the non-commutativity between the operators $H_{n,l}$, $H_{m,l}$ and $H_{n,m}$. Higher accuracy is realised by considering a smaller time step. Here, we took $\Delta t = 0.1$ fs. Using the well-known property of the Pauli matrices
\begin{equation}
\exp\left[-i\vec{A}\cdot\vec{\sigma}\right] = \cos(A)\textbf{I} - i \frac{\sin(A)}{A} \left(\begin{matrix}
A_z & A_x-iA_y \\
A_x+iA_y & -A_z \\
\end{matrix}\right)
,\end{equation}
for any vector $\vec{A}$, where $A = |\vec{A}|$ and $\textbf{I}$ is the identity matrix, and realising that the $H_{n,m}$ operator for each $n$ and $m$ fixed is just a $2\times 2$ matrix with zero-diagonal elements described by $\Delta_{n,m}\sigma_x$, we have that the exponential of $H_{n,m}$ is given exactly by
\begin{equation}
e^{-(i/2\hbar)\Delta_{n,m}\sigma_x \Delta t} = \left(\begin{matrix}
\cos(A_x) & - i \sin(A_x) \\
- i \sin(A_x) & \cos(A_x) \\
\end{matrix}\right) = \mathcal{M}_l
,\end{equation}
where $A_x = \Delta_{n,m}\Delta t/2\hbar$.

The wave function at time step $t + \Delta t$ is then given by
\begin{align}
|\Psi_{n,m,l}\rangle_{t + \Delta t} &\cong e^{-(i/2\hbar)H_{n,m} \Delta t} e^{-(i/2\hbar)H_{m,l} \Delta t} e^{-(i/\hbar)H_{n,l}\Delta t}\nonumber\\
&\times e^{-(i/2\hbar)H_{m,l} \Delta t} e^{-(i/2\hbar)H_{n,m} \Delta t}|\Psi_{n,m}\rangle_{t}
,\end{align}
that can be developed in five steps
\begin{equation}
\eta_{n,m,l} = e^{-(i/2\hbar)H_{n,m} \Delta t}|\Psi_{n,m,l}\rangle_{t}
,\end{equation}
\begin{equation}
\xi_{n,m,l} = e^{-(i/2\hbar)H_{m,l} \Delta t}\eta_{n,m,l}
,\end{equation}
\begin{equation}
\chi_{n,m,l} = e^{-(i/\hbar)H_{n,l}\Delta t}\xi_{n,m,l}
,\end{equation}
\begin{equation}
\varrho_{n,m,l} = e^{-(i/2\hbar)H_{m,l} \Delta t}\chi_{n,m,l}
,\end{equation}
\begin{equation}
|\Psi_{n,m,l}\rangle_{t + \Delta t} = e^{-(i/2\hbar)H_{n,m} \Delta t} \varrho_{n,m,l}
,\end{equation}
where at each step we use the Cayley equation for the exponentials \cite{Watanabe}, such that
\begin{equation}
\eta_{n,m,l} = \mathcal{M}_l |\Psi_{n,m,l}\rangle_{t}\nonumber
,\end{equation}
\begin{equation}
\left(1 + \frac{i\Delta t}{4\hbar}H_{m,l}\right) \xi_{n,m,l} = \left(1 - \frac{i\Delta t}{4\hbar}H_{m,l}\right) \eta_{n,m,l}\nonumber
,\end{equation}
\begin{equation}
\left(1 + \frac{i\Delta t}{2\hbar}H_{n,l}\right) \chi_{n,m,l} = \left(1 - \frac{i\Delta t}{2\hbar}H_{n,l}\right) \xi_{n,m,l}\nonumber
,\end{equation}
\begin{equation}
\left(1 + \frac{i\Delta t}{4\hbar}H_{m,l}\right) \varrho_{n,m,l} = \left(1 - \frac{i\Delta t}{4\hbar}H_{m,l}\right) \chi_{n,m,l}\nonumber
,\end{equation}
\begin{equation}
|\Psi_{n,m,l}\rangle_{t + \Delta t} = \mathcal{M}_l \varrho_{n,m,l}
.\end{equation}
The problem is now strongly simplified because now we have to deal only with tridiagonal matrices. We propagate a Gaussian wave packet following this numerical procedure and calculate the transmission probability by integrating the squared modulus of the wave packet only in the region of the bilayer after the QPC. The initial Gaussian wave packet is defined as:
\begin{align}
\Psi_0(\vec{r}) =& \frac{1}{d\sqrt{2\pi}}
\left(\begin{array}{c}
A\\
B\\
A'\\
B'
\end{array}\right)\nonumber\\
& \times\exp\left[{-\frac{(x-x_0)^2+(y-y_0)^2}{2d^2}+i\vec{k}\cdot\vec{r}}\right]
.\end{align}
The coefficients $A(A')$ and $B(B')$ in the pseudospinor are related to the probability of finding the electron in each triangular sublattice $A(A')$ and $B(B')$ of the graphene lattice in a given layer. For the bilayer case, we choose the same pseudospinor for both layers, since the total wave function is composed of two Gaussian wave packets, one in each layer, with the same properties, as initial momentum, initial energy and initial position of Gaussian center $\vec{r}_0 = (x_0,y_0)$ in real space. The pseudospinor is characterised by the pseudospin polarization angle $\theta$, such as $\left(1 \mbox{~,~} e^{i\theta}\right)^T$. Thus the pseudospin polarization has a conceptual connection with the direction of propagation of the wave packet in the tight-binding model and the choice of the angle $\theta$ depends also on which Dirac valley the initial wave packet is taken\cite{Andrey3, Diego}. We take $\theta = 0$($\pi$) for an initial wave packet starting from $K$($K'$) valley, since we want it to propagate in the $y$-direction. The initial wave vector is $\vec{k} = (k^0_x,k^0_y) + \overline{K}$, which is shifted with respect to the Dirac points, where $\overline{K}$ represents the two non-equivalent $K$ and $K'$ points that are located at $(0,\pm 4\pi/3\sqrt{3a})$, with $a=0.142$ nm being the in plane inter-atomic distance. For our numerical calculations, the initial wave packet energy $E$ is set by the modulus of the wave vector $k$, since the bottom of the low-energy bands may be approximated by\cite{KoshinoReview} $E=-(\tau_{\bot}/2)\left(\sqrt{1 + 4 (v_F\hbar k/\tau_{\bot})^2}-1\right)$, in bilayer graphene, where $\tau_{\bot} \approx 0.4$ eV is the interlayer coupling corresponding to perpendicular hopping between the Bernal stacked layers and $v_F \approx 10^6$ m$/$s is the Fermi velocity. The width of the Gaussian wave packet was taken as $d = 20$ nm and its initial position as $(x_0,y_0) = (0,-42)$ nm. If the wave packet width in ($x$-) $k$-space is (small) large, it will be composed of a distribution of plane-waves with different velocities and, therefore, exhibit a strong decay in time, due to the parabolic dispersion in $k$-space. We have checked that the wave packet width in real space considered in our calculations is appropriate for the proposed problem, being large enough to avoid significant changes of the wave packet within the time scale of interest.

An important remark concerning the wave packet dynamics is about the oscillatory behavior of the velocity, \textit{i.e}. the zitterbewegung manifestation on the wave packet motion \cite{Andrey3, Maksimova}. We shall show that it can not be avoided for motion of an electron in bilayer graphene that propagates in the $y$-direction. To understand how this affects the velocity in the $y$-direction, we use the Dirac Hamiltonian for electrons in bilayer graphene in the vicinity of the $K$ point\cite{KoshinoReview}
\begin{equation}\label{eq.HD}
H^{BI}_D = \left(\begin{tabular}{cccc}
0 & $v_F \pi$ & $\tau_{\bot}$ & $0$\\
$ v_F \pi^{\dagger}$ & 0 & $0$ & $0$\\
$\tau_{\bot}$ & $0$ & 0 & $v_F \pi^{\dagger}$\\
$0$ & $0$ & $v_F \pi$ & 0\end{tabular} \right),
\end{equation}
where $\pi = p_x + ip_y$ and $\pi^{\dagger} = p_x - ip_y$ are the momentum operators in Cartesian coordinates, and calculate the commutator $[H^{BI}_D, v^{BI}_y]$. According to the Heisenberg picture, the velocity in the $y$-direction is given by
\begin{equation}\label{eq.vy.bi}
v_y = \frac{d y}{dt} = \frac{1}{i\hbar}[H_D, y]
.\end{equation}
Replacing $H^{BI}_D$ into Eq. (\ref{eq.vy.bi}) and using the well-known commutation relation $[x_i,p_j] = i\hbar\delta_{ij}$ we find
\begin{equation}\label{eq.vy.bi.final}
v^{BI}_y = \left(\begin{tabular}{cccc}
$0$ & $iv_F$ & $0$ & $0$\\
$-iv_F$ & $0$ & $0$ & $0$\\
$0$ & $0$ & $0$ & $-iv_F$\\
$0$ & $0$ & $iv_F$ & $0$ \end{tabular} \right).
\end{equation}

Now we shall verify whether $v^{BI}_y$ is a constant of motion or not, and if there is any situation where the velocity is not affected by the zitterbewegung in the $y$-direction. Evaluating $[H^{BI}_D, v^{BI}_y]$ by making use of Eqs. (\ref{eq.HD}) and (\ref{eq.vy.bi.final}), one obtains
\begin{equation}\label{eq.commutator.bi}
[H^{BI}_D, v^{BI}_y] = \left(\begin{tabular}{cccc}
$-2iv^2_Fp_x$ & $0$ & $0$ & $-i\tau_{\bot}v_F$\\
$0$ & $2iv^2_Fp_x$ & $i\tau_{\bot}v_F$ & $0$\\
$0$ & $i\tau_{\bot}v_F$ & $2iv^2_Fp_x$ & $0$\\
$-i\tau_{\bot}v_F$ & $0$ & $0$ & $-2iv^2_Fp_x$ \end{tabular} \right),
\end{equation}
suggesting that even if $p_x = 0$, one has $[H^{BI}_D, v^{BI}_y]\neq 0$, implying that $v_y$ is not a constant of motion, because we are still left with non-zero off-diagonal terms. Conversely, in the monolayer case, we obtain that the velocity in the $y$-direction is expressed by $v^{MO}_y = -v_F\sigma_y$, where $\sigma_y$ is the $y$ Pauli matrix and with the monolayer Hamiltonian being $H^{MO}_D = v_F\vec{\sigma} \cdot\vec{p}$. Following the same procedure as for bilayer graphene, we obtain
\begin{equation}\label{eq.commutator.mono}
[H^{MO}_D, v^{MO}_y] = \left(\begin{tabular}{cc}
$-2iv^2_Fp_x$ & $0$\\
$0$ & $2iv^2_Fp_x$\end{tabular} \right),
\end{equation}
and thus $[H^{MO}_D, v^{MO}_y] = 0$ if $p_x = 0$. Therefore, there is no way to avoid the trembling motion at small times for the wave packet propagation in bilayer graphene - even for motion in the $y$-direction, \textit{i. e.} $k^0_x = 0$ and $k^0_y\neq 0$, the wave packet will also move in the $x$-direction. For this reason, all presented results in this paper are normalized to the maximum transmission obtained for the case in the absence of any potential.

\section{Results and discussion}\label{sec.results}

\begin{figure}[!bpht]
\centerline{\includegraphics[width = \linewidth]{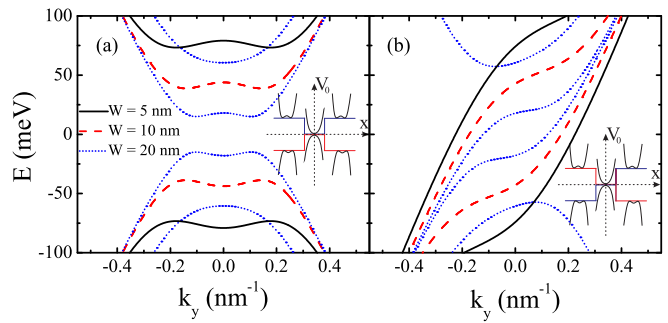}}
\caption{(Color online) Band structure for the two potential configurations sketched in Fig. \ref{fig:Fig1}, namely, (a) aligned and (b) anti-aligned potential barriers. Results are presented for $V_0 = 200$ meV and three values for the quantum well width $W = 5$ (black solid), $10$ (red dashed) and $20$ nm, (blue dotted).} \label{fig:Fig2}
\end{figure}

In order to better understand our tight-binding results, let us first investigate the energy dispersions in a BLG quantum wire defined by aligned (Fig. \ref{fig:Fig1}(a)) and anti-aligned (Fig. \ref{fig:Fig1}(b)) potential barriers, as obtained by the Dirac approximation for BLG, using the 4$\times$4 Hamiltonian. \cite{KoshinoReview} These spectra are shown in Figs. \ref{fig:Fig2}(a) and \ref{fig:Fig2}(b), respectively, for different values of well width. Only states with energy below the barrier height ($V_0 = 200$ meV) are shown. Notice the qualitative difference between the two spectra. The energy dispersion in the former case exhibits symmetry with respect to positive and negative values of the wave vector in the propagation direction $k_y$. As the propagation velocity is obtained from $v_g = \partial E/ \partial p_y$, with $p_y = \hbar k_y$, this spectrum suggests that wave packets may propagate towards either positive or negative $y$-direction, provided the average wave vector $k_y$ of the wave packet is in a region of positive or negative derivative of the spectrum, respectively. On the other hand, the spectra for the anti-aligned case does not exhibit the same symmetry. Besides, low energy electrons in this system can only exhibit positive velocity of propagation, since the derivative of the spectrum around $E = 0$ is positive for any value of $k_y$. In fact, this spectrum is obtained for the BLG Dirac Hamiltonian for electrons around the $K$ point of the first Brillouin zone. The spectrum for $K'$ is obtained just by replacing $k_y$ by $-k_y$ in Fig. \ref{fig:Fig2}(b), or, equivalently by inverting the polarization of all gates in Fig. \ref{fig:Fig1}(b). Thus, although the band structures in $K$ and $K'$ valleys are different, they are still mirror-matched.

\begin{figure}[b]
\centerline{\includegraphics[width = 0.8\linewidth]{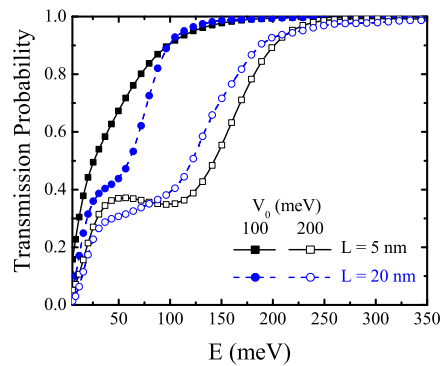}}
\caption{(Color online) Transmission probability as a function of wave packet energy for aligned potentials with $V_0 = 100$ meV (closed symbols) and $V_0 = 200$ meV (open symbols). The width of the QPC is $W = 10$ nm. The square and circular symbols correspond to the QPC lengths $L = 5$ and $20$ nm, respectively.} \label{fig:Fig3}
\end{figure}

Due to the fact that the low energy spectra for the aligned bias case in $K$ and $K'$ points have the $E_{K(K')}(k_x, k_y) = E_{K(K')}(-k_x, k_y)$ and $E_{K}(k_x, k_y) = E_{K'}(k_x, -k_y)$ symmetries, the transmission probabilities are the same no matter if the wave packet started in $K$ or $K'$ valley. This is verified in Fig. \ref{fig:Fig3}, which shows the transmission probability as a function of the wave packet energy, for different configurations of the channel defined by aligned potentials, as sketched in Fig. \ref{fig:Fig1}(a). Results for $K$ and $K'$ valleys in this case are exactly the same. Steps are observed in the transmission probabilities as the wave packet energy increases. This is a well known feature of any QPC, which is related to the existence of quantized energy levels inside the channel - whenever the energy crosses one of the energy levels, a step is produced. In fact, although not shown in Fig. \ref{fig:Fig3}, we verified that increasing $W$ moves the steps to lower $E$, just as expected for an usual QPC, once that the energy of the quantized states of the channel decreases as $W$ increases. The figure shows that as $L$ varies the position of the step does not change, but for smaller $L$, the steps are less pronounced.

\begin{figure}[!bpht]
\centerline{\includegraphics[width = \linewidth]{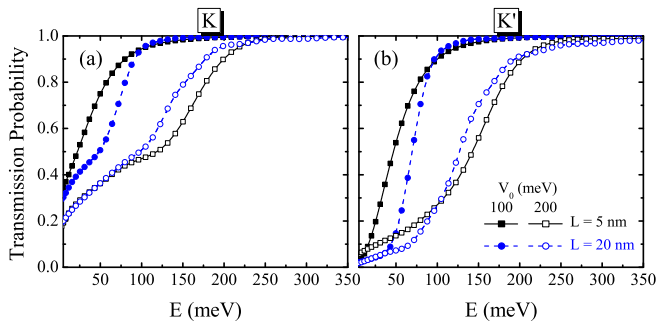}}
\caption{(Color online) The same as Fig. \ref{fig:Fig3}, but for anti-aligned potentials and with initial wave packet in $K$ valley (a) and in $K'$ valley (b).} \label{fig:Fig4}
\end{figure}

Similar features are observed in the anti-aligned case, though with a fundamental difference - results for $K$ and $K'$ in this case are very different, as one can verify by comparing Figs. \ref{fig:Fig4}(a) and \ref{fig:Fig4}(b), respectively. This is a clear manifestation of the lack of inter-valley symmetry exhibited by the band structure shown in Fig. \ref{fig:Fig2}(b). Such a difference between transmission probabilities in different valleys suggests the use of this system as a valley filter. However, in order to do so, we should seek for the best configuration of the system that enhances valley polarization.

\begin{figure}[b]
\centerline{\includegraphics[width = \linewidth]{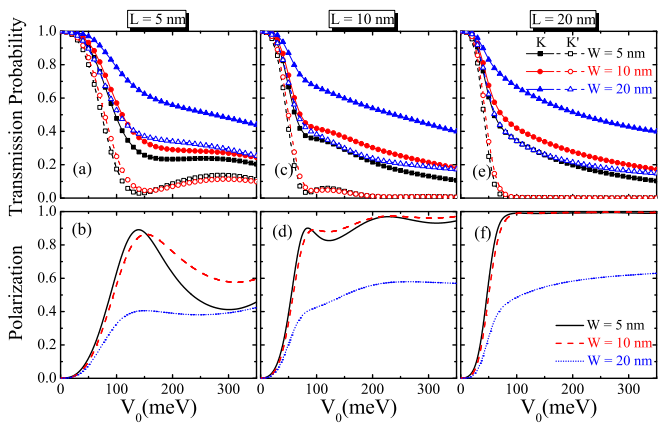}}
\caption{(Color online) Transmission probability (top panels) and valley polarization (bottom panels) as a function of the electrostatic bias $V_0$ in the case of anti-aligned potentials with initial wave packet energy $E = 30$ meV and three different values of $L$: (a, b) $5$ nm, (c, d) $10$ nm and (e, f) $20$ nm. The opened (closed) square-like, circular and triangular symbols correspond to $W = 5$, $10$ and $20$ nm, for the $K$ ($K'$) valley, respectively in panels (a), (c) and (e). The black solid, red dashed and blue dotted lines show the polarization for $W = 5$, $10$ and $20$ nm, respectively in panels (b), (d) and (f).} \label{fig:Fig5}
\end{figure}

Figure \ref{fig:Fig5} shows the transmission probabilities (upper panels) and the valley polarization (lower panels) for the anti-aligned system of Fig. \ref{fig:Fig1}(b) as a function of the bias potential $V_0$, for different values of $W$ and $L$, considering a wave packet energy $E = 30$ meV. Once the wave packet is initially injected in the lowest subband of the energy spectrum (Fig. \ref{fig:Fig2}), then we limit ourselves to the lowest QPC steps. Valley polarization is defined as $P = 1-T_{K'}/T_K$, where $T_{K(K')}$ is the transmission probability for a wave packet starting at the $K (K')$ Dirac points, so that $P = 1$ (0) means a wave packet completely (un)polarized in $K$ after the QPC. Transmission probabilities in all cases are reduced as $V_0$ increases, which is expected, since the existence of a barrier leads to stronger reflection of the tails of the wave packet that are outside the channel region. A very weak oscillation is observed in each curve, which is due to an interference related to the path difference between electrons that go straight through the channel and those that are reflected at the exit and entrance of the channel. Results in Figs. \ref{fig:Fig5} (b, d, f) show a polarization that increases up to $1$ for higher values of $L$ and $V_0$, in particular for small $W$. For larger $W$, however, the electron starts to see a larger unbiased area in the channel, thus reducing the polarization effect. This polarization reduction for large $W$ becomes even more significant for wave packets with higher energy. Indeed, extra energy bands with higher energy appear as $W$ increases (see Fig. \ref{fig:Fig1}(b), blue dotted curves). These bands exhibit states with negative velocities, which, consequently, harness the polarization effect proposed here, which relies on bands with a single direction of the propagation velocity. The results in Fig. \ref{fig:Fig4} demonstrate that as the initial energy of the wave packet $E$ increases, higher transmission probabilities are reached for wave packets starting in both valleys ($K$ and $K'$) and thus a suppression of the polarization effect is expected in this case, which is due to the low screening of the packet by the barriers for a fixed range of bias potential. It is also clear that increasing the channel length improves the valley polarization.

\begin{figure}[!bpht]
\centerline{\includegraphics[width = \linewidth]{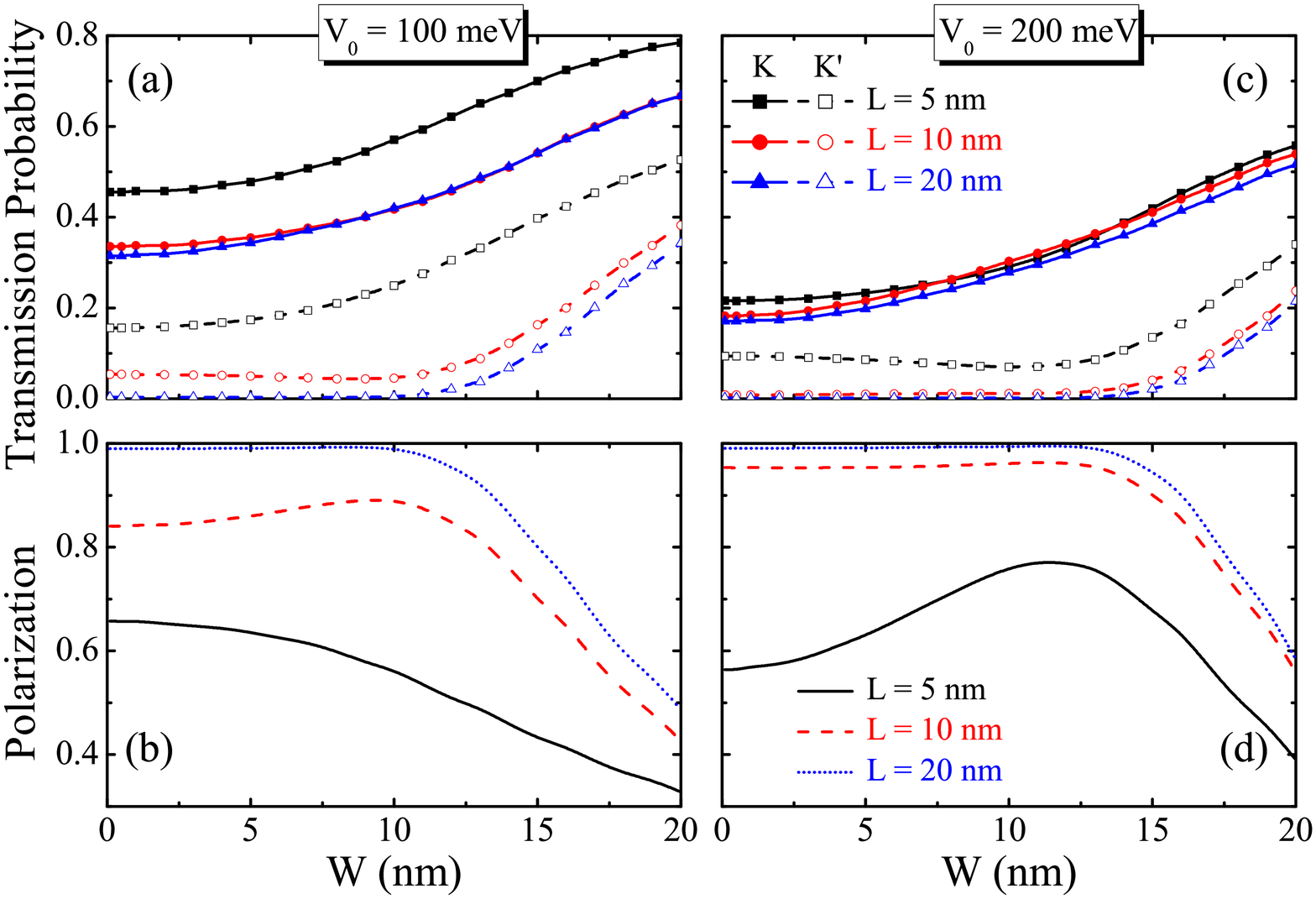}}
\caption{(Color online) (a, c) Transmission probability and (b, d) the polarization as a function of the width $W$ of the QPC for anti-aligned potentials with $V_0 = 100$ meV (left side panels) and $V_0 = 200$ meV (right side panels). The average wave packet energy was $E = 30$ meV. The opened (closed) square-like, circular and triangular symbols correspond to $L = 5$, $10$ and $20$ nm, for the $K$ ($K'$) valley, respectively in Figs. (a) and (c). The black solid, red dashed and blue dotted curves show the polarization for $L = 5$, $10$ and $20$ nm, respectively in Figs. (b) and (d).} \label{fig:Fig6}
\end{figure}

As already mentioned, even such almost perfect polarization for large $L$ can be destroyed by increasing $W$. This is clarified in Fig. \ref{fig:Fig6}, which shows transmission probabilities (upper panels) and polarization (lower panels) as a function of the well width $W$ in an anti-aligned QPC. In the case of $L = 20$ nm (dotted blue line), polarization stays around $\approx 100 \%$ for smaller $W$, but starts to decrease for $W > 10$ nm and $13$ nm, in the cases of $V_0 = 100$ meV and $200$ meV, respectively.

Although all the presented results in this paper were obtained for a circular Gaussian wave packet that propagates straight through the gated constriction and whose center of mass position is located initially at the middle of the $x$-axis, it is worth to mention about the robustness of the polarization results when these conditions are not met. Results for an elliptic Gaussian wave packet, with larger width in the $x$-direction, as well as for wave packets propagating with non-zero angle with respect to the $y$-axis, have been obtained too, although only discussed here in a qualitative way in what follows. The elliptic case shows quite similar qualitative features, so that the polarizations are kept practically the same, especially in the range of most efficient polarization. For instance, considering $V_0 = 0.2$ eV, $W = 5$ nm and $E = 30$ meV, we found that making the wave packet $50 \%$ larger in the $x$-direction would lead to a maximum difference of $\approx 13.5\%$ for intermediate values of polarization, whereas in the range of most efficient polarization, it shows no significant change. For a much larger channel width $W = 20$ nm and considering such larger wave packet, changes in polarization are not larger than $\approx 3.6 \%$. On the other hand, the transmission probabilities for each valley decreases, since a larger part of the wave packet is reflected by the biased (gapped) regions of the system in this case. Also, the packets along different propagation angles or different axis would, of course, be largely reflected by the potential barriers. The filtering efficiency, however, depends only on the ratio between transmission probabilities in $K$ and $K'$ valleys; details of the wave packet, such as its width, would not significantly modify this result, since they would change the transmission probabilities for both $K$ and $K'$ almost in the same way, keeping the ratio and, consequently, the filtering efficiency, practically unchanged.

In summary, our results demonstrate that an almost perfect valley filtering can be realised, provided (i) the electron energy is sufficiently low, (ii) the channel length is sufficiently long, and (iii) the channel width is narrow enough.

\section{Conclusion}\label{conclusion}

We calculated the transmission probabilities of a Gaussian wave packet through a quantum point contact defined by electrostatic gates in bilayer graphene. Our results demonstrate that, if one uses the energy gaps introduced by a bias between upper and lower layers in order to define the channel in the point contact, transmission plateaus are observed as the energy of the packet increases, which reflects the discrete eigenstate spectrum in the channel, just like in a conventional QPC. On the other hand, if the bias in the left and right sides of the channel are opposite to each other, although still forming the same energy gap at both sides, a special situation of energy dispersion is obtained, where electrons in each valley have only one possible direction of propagation. In this case, the QPC works as an efficient valley filter, where valley polarization may reach $\approx 1$ with increasing gate potential. Such a valley filtering device can have an important impact on future graphene valley-tronics, as it can be relatively easily achieved just by depositing electrostatic gates on graphene\cite{Chua}, with no need either to control edge types, or to produce strain or non-zero mass regions, in contrast to the valley filters previously proposed in the literature.

\acknowledgements This work was financially supported by CNPq, under the PNPD and the PRONEX/FUNCAP grants, CAPES Foundation under the process number BEX 7178/13-1, the Bilateral programme between Flanders and Brazil, the Flemish Science Foundation (FWO-Vl), and the Brazilian Program Science Without Borders (CsF).

\end{document}